\documentclass[10pt,showpacs,preprintnumbers,amsmath,amssymb,floatfix,prd]{revtex4}
\usepackage[dvips]{graphicx}
\begin{document}

\title{The three flavor LOFF phase of QCD }

\author
{Nicola~D.~Ippolito}\email{nicola.ippolito@ba.infn.it}
\date{\today}
\affiliation{Universit\`a di Bari, I-70126 Bari, Italy}
\affiliation{I.N.F.N., Sezione di Bari, I-70126 Bari, Italy}
\preprint{BARI-TH/06-552}

\begin{abstract}
 We explore the
Larkin-Ovchinnikov-Fulde-Ferrell (LOFF) phase
of QCD with three flavors, using a
Ginzburg-Landau expansion of the free energy,
and  a NJL point-like four-fermion
interaction, with the quantum numbers of
single-gluon exchange.
\end{abstract}

\maketitle

 \vskip 10mm

At high quark densities and small
temperatures quarks are expected to form
Cooper pairs, because of the attractive
interaction in the color antisymmetric
channel. This gives rise to Color
Superconductivity,
see~\cite{barrois,Alford:1997zt} and for
reviews \cite{review,Nardulli:2002ma}. At
asymptotically high densities with three
flavors the Color Flavor Locking (CFL)
phase~\cite{Alford:1998mk}, characterized by
a spinless color- and flavor-antisymmetric
diquark condensate, is the energetically
favored phase. When the density decreases to
values probably achieved in the core of
compact stars ($\mu \sim 400-500$ MeV) one
cannot neglect the strange quark mass and the
differences in the quark chemical potentials
$\delta \mu$. Various phases have been
proposed for this region of the QCD phase
diagram. Among these, the gapless gCFL phase
\cite{Alford:2003fq} was considered for some
time as the most suitable candidate. It was
however realized that imaginary gluon
Meissner masses,
see~\cite{Casalbuoni:2004tb}, produce
instabilities in this phase, and in general in the gapless phases, see~\cite{Huang:2004bg}. 

The Larkin-Ovchinnikov-Fulde-Ferrell (LOFF)
\cite{LOFF2} phase is an inhomogeneous
superconductive phase in which quark pairs
have non-zero total momentum; it turns out to
be favored, for appropriate values of $\delta
\mu$, over the homogeneous superconducting
phase and the normal non-superconducting one.
The possibility of a LOFF phase in QCD with
two flavors has been extensively discussed in
the last half-decade,
see~\cite{Alford:2000ze} and for a
review~\cite{Casalbuoni:2003wh}.

The three flavor LOFF phase has been first
considered in \cite{Casalbuoni:2005zp}, using
a Ginzburg-Landau (GL) expansion of the free
energy, and subsequently it was found that
such a phase is chromomagnetically stable
\cite{Ciminale:2006sm}. The validity of the
G-L approximation for this phase has recently
been tested \cite{Mannarelli:2006fy}, and a
study of the corrections due to the finite
chemical potentials has been completed in
~\cite{Casalbuoni:2006zs}. Recently a work
has been performed  \cite{Rajagopal:2006ig}
in which the authors consider the possibility
of structures formed by more than one plane
wave.

\begin{center}
\section*{Formalism and results}
\end{center}

We start with  the Lagrangean density for
three flavor ungapped quarks
\begin{equation}
{\cal
L}=\bar{\psi}_{i\alpha}\,\left(i\,D\!\!\!\!
/^{\,\,\alpha\beta}_{\,\,ij}
-M_{ij}^{\alpha\beta}+ \mu^{\alpha\beta}_{ij}
\,\gamma_0\right)\,\psi_{\beta j}
\label{lagr1}\ .
\end{equation}
 $M_{ij}^{\alpha\beta} =\delta^{\alpha\beta}\, {\rm diag}(0,0,M_s)
$ is the mass matrix and
$D^{\alpha\beta}_{ij}=\partial\delta^{\alpha\beta}\delta_{ij}+
igA_aT_a^{\alpha\beta}\delta_{ij}$;
$\mu_{\alpha\beta}^{ij}$ is a diagonal
color-flavor matrix depending in general on
$\mu$ (the average quark chemical potential),
$\mu_e$ (the electron chemical potential),
and $\mu_3,\,\mu_8$, related to color
\cite{Alford:2003fq}. We do not require color
neutrality and we work in the approximation
$\mu_3=\mu_8=0$, which is justified by the
results of Ref.\cite{Casalbuoni:2006zs}, so
we write
$\mu^{\alpha\beta}_{ij}=(\mu\delta_{ij}-\mu_e
Q_{ij})\delta^{\alpha\beta}=\mu_{i}\,\delta_{ij}\delta^{\alpha\beta}$.\\
Treating the strange quark mass at the
leading order in the $1/\mu$ expansion we
have
  \begin{equation}
\mu_u=\mu-\frac{2}{3}\mu_e~,\
\mu_d=\mu+\frac{1}{3}\mu_e~,\
\mu_s=\mu+\frac{1}{3}\mu_e-\frac{M_s^2}{2\mu}~.\label{eq:defChemPotQuarks}\end{equation}
Moreover we make use of the  High Density
Effective Theory (HDET), see
\cite{Hong:1998tn} and, for a review,
\cite{Nardulli:2002ma}.

The fundamental pairing ansatz for the condensate is
\begin{equation}
<\psi_{i\alpha}\,C\,\gamma_5\,\psi_{\beta j}> =
\sum_{I=1}^{3}\,\Delta_I({\bf r})\,\epsilon^{\alpha\beta
I}\,\epsilon_{ijI}~\label{cond}
\end{equation}with \begin{equation} \Delta_I ({\bf r}) = \Delta_I
\exp\left(2i{\bf q_I}\cdot{\bf r}\right)\,, \label{eq:1Ws}\end{equation}
where $2{\bf q_I}$ is
the pair momentum. The momenta ${\bf q_1}$, ${\bf q_2}$, ${\bf q_3}$ and the gap parameters
$\Delta_{1}({\bf r})$, $\Delta_{2}({\bf r})$, $\Delta_{3}({\bf r})$
correspond respectively to $d-s$, $u-s$ and $u-d$ pairing.

As usual, the norms $|{\bf q_I}|$ are derived minimizing the free energy, while
the directions have to be determined by a chrystallographic analysis. 
In \cite{Casalbuoni:2005zp} only structures
with the $\bf q_I$ on the same axis have been
studied, while in \cite{Rajagopal:2006ig} more complicate structures are analyzed.

The GL approximation is implemented by an expansion of the anomalous propagator $S_{21}$
\begin{equation}
S_{21}=S_0^{22}\Delta^*S_0^{11} +S_0^{22}\Delta^*S_0^{11}\Delta
S_0^{22}\Delta^*S_0^{11} +O(\Delta^5)\, \label{eq:GLexpansion} \end{equation}
involved in the gap equation, as obtained by the Schwinger-Dyson equation in
the HDET formalism:
\begin{equation}
\Delta^*_{AB}({\bf r})
=\,i\,3G\,V^\mu\tilde{V}^\nu
\sum_{C,D=1}^9\!h^*_{AaC}h_{DbB}\int\frac{d\,{\bf
n}}{4\pi} \int\frac{d^3\,\ell}{(2\pi)^3}
\int\frac{dE}{2\pi}\,S_{21}(E,\ell)_{CD}\,g_{\mu\nu}\,\delta_{ab}~.
\label{eq:GapEq122}
\end{equation}
In this way the free energy GL expansion assumes the form
\begin{equation}
\Omega =\Omega_n+ \sum_{I=1}^3\left(\frac{\alpha_I}{2}\,\Delta_I^2
~+~ \frac{\beta_I}{4}\,\Delta_I^4 ~+~ \sum_{J\neq
I}\frac{\beta_{IJ}}{4}\,\Delta_I^2\Delta_J^2 \right) ~+~ O(\Delta^6)
\label{eq:OmegaDelta}
\end{equation}
with \begin{equation}
 \Omega_n = -\frac{3}{12\pi^2}\left(\mu_u^4+\mu_d^4+\mu_s^4\right) -
\frac{\mu_e^4}{12\pi^2} ~. \label{eq:OmegaNorm11222}
\end{equation}
The coefficients $\alpha_I$, $\beta_I$ and $\beta_{IJ}$ can be found
in \cite{Casalbuoni:2005zp}.
Electric neutrality is obtained by imposing the condition
\begin{equation}
-\frac{\partial\Omega}{\partial\mu_e}=0~. \label{eq:electrNeutra111}
\end{equation}.

\begin{figure}[h!]
\begin{center}
{\includegraphics[width=10cm]{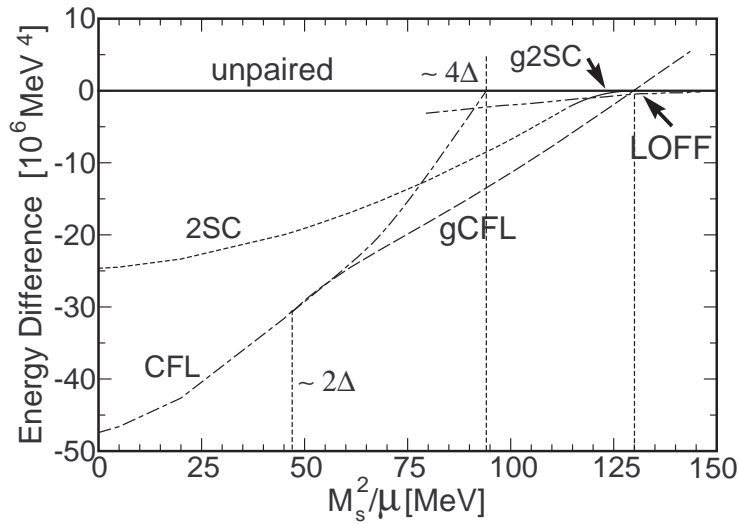}}
\end{center}
\caption{\label{FIG:comparison500} \footnotesize{ Free energy differences
$\Omega_{LOFF}-\Omega_{norm}$ in units of $10^{6}$ MeV$^4$ plotted
versus  $M_s^2/\mu$ (in MeV)  for various QCD phases.}}
\end{figure}

In \cite{Casalbuoni:2005zp} the energetically favored solution has $\Delta_1=0$,
$\Delta_2=\Delta_3$ and $\bf q_2$, $\bf q_3$ parallel.
Figure \ref{FIG:comparison500} shows the competition among
different color superconducting phases, comprising 
the three flavor LOFF phase considered in \cite{Casalbuoni:2005zp}.

In Ref.\cite{Rajagopal:2006ig} another plot
is presented, showing the free energies of
more complicate crystal structures. In
particular, there are two structures with
free energy smaller than the simple ansatz of
\cite{Casalbuoni:2005zp}, but the gaps
involved in this phases are rather large, so
the GL approximation must be properly tested.

In conclusion, we have shown that the three
flavor LOFF phase, due to its chromomagnetic
stability, is a serious candidate for the
true vacuum at moderate densities, and
therefore it could be present in the core of
compact stars (see \cite{Anglani:2006br} for
a preliminary study).

\vspace{1cm}

{\em Acknowledgments}

I wish to thank G. Nardulli and M. Ruggieri,
for their fundamental importance upstream
this work; D. Blaschke and BLTP-JINR for
everything concerning the DM 2006 school; R.
Anglani and all the other participants to DM
2006 for two really useful and funny weeks.

\end{document}